\documentclass[final,preprint,draft,showpacs,tightenlines,aps]{revtex4}%
\usepackage{amsmath}
\usepackage{epsfig}
\usepackage{graphicx}%
\usepackage{amsfonts}%
\usepackage{amssymb}

\begin{document}

\begin{flushright}
TPJU-11/2003 \\
BNL-NT-03/38
\end{flushright}
  \vspace*{1cm}
\title{The width of $\Theta^{+}$ for large $N_{c}$ in chiral quark soliton model}
\author{Micha{\l} Prasza{\l}owicz\footnote{E-mail address:~michal@quark.phy.bnl.gov,
Fulbright Fellow on leave from M. Smoluchowski Institute of Physics,
Jagellonian University, Krak{\'o}w, Poland}}
\affiliation{Nuclear Theory Group, \\
Brookhaven National Laboratory, \\
Upton, NY 19973-5000 \\
~~~}

\begin{abstract}
In the chiral quark soliton model the smallness of $\Theta^+$
width is due to the cancellation of the coupling constants which
are of different order in $N_c$. We show that taking properly into
account the flavor structure of relevant SU(3) representations for
arbitrary number of colors enhances the nonleading term by an
additional factor of $N_c$, making the cancellation consistent
with the $N_c$ counting. Moreover, we show that, for the same
reason, $\Theta^+$ width is suppressed by a group-theoretical
factor ${\cal O}(1/N_c)$ with respect to $\Delta$ and discuss the
$N_c$ dependence of the phase space factors for these two decays.
\end{abstract}

\maketitle

%\date{November 2003}

\section{Introduction}

Recently five experiments announced discovery of a narrow, exotic baryonic
state called $\Theta^{+}$ \cite{exp}. Most probably this state belongs to the
positive parity baryon antidecuplet, which naturally emerges in chiral soliton
models \cite{manchem,prasz,DPP}.
Early prediction of its mass \cite{prasz} obtained in the Skyrme model
\cite{Skyrme} extended to three flavors \cite{SU3SM} is in
surprising agreement with the present experimental findings. Moreover, if the
discovery of the heaviest members of the antidecuplet, $\Xi_{3/2}$, announced
by NA49 collaboration \cite{NA49} at 1860 MeV is confirmed,
again the same model will be off only by 70 MeV \cite{prasz}.

However, the most striking experimental result is perhaps
a very small width $\Gamma_{\Theta^{+}}$ which is estimated to be of the order
of a few tens MeV or less \cite{exp}. Such a narrow width was predicted in a seminal
paper by Diakonov, Petrov and Polyakov \cite{DPP} within the SU(3) chiral quark
soliton model ($\chi$QSM). The smallness of $\Gamma_{\Theta^{+}}$ in this model
is due
to the cancellation of the coupling constants which enter the collective decay
operator. It is, however, at first sight to some extent unnatural that the
cancellation occurs between the constants which are of different order in the
number of colors, $N_{c}$. If only the leading term were retained, the width
of $\Theta^+$ would be of the same order as $\Gamma_{\Delta}$ \cite{DPP,Weigel}.

In this note we show that in fact this cancellation occurs order
by order in $N_{c}$. This is due to the fact that the additional
factor of $N_{c}$ appears when one properly takes into account the
SU(3) flavor representation content of the lowest lying baryonic
states. Indeed, for arbitrary number of colors, baryons do not
fall into ordinary flavor octet, decuplet and antidecuplet, but
are members of large SU(3) representations
\cite{largereps,dulmp,dul}, which reduce to the standard ones only
for $N_{c}=3$. Taking this extra $N_{c}$ dependence into account
makes the SU(3) Clebsch-Gordan coefficients depend on $N_{c}$.

We also find that group-theoretical factors suppress
$\Gamma_{\Theta^+}$ with respect to $\Gamma_{\Delta}$. This
suppression is, however, "undone" by the phase space factor, which
scales differently with  $N_c$ in the chiral limit.

\section{Large $N_{c}$ limit}

Both in the Skyrme model \cite{Skyrme,SU3SM} and in the $\chi$QSM
\cite{inst,Blotz0} baryons emerge as rotational
states of the symmetric top which rotates in the SU(3) collective space
\cite{prasz,SU3SM}. This
rotation is described in terms of a rotational SU(3) matrix $A(t)$. Baryonic
wave functions \cite{SU3SM,DPP,Blotz0,Blotz} are given as SU(3) Wigner
$D_{BJ}^{(\mathcal{R})}(A)$ matrices~\cite{foot1}:
\begin{equation}
\psi_{BS}^{(\mathcal{R})}(A)=\psi_{(\mathcal{R},B)(\mathcal{R}^{\ast}%
,S)}(A)=(-)^{Q_{S}}\sqrt{\dim(\mathcal{R})}D_{BJ}^{(\mathcal{R})\ast}(A).
\label{wf}%
\end{equation}
In this notation $B=(Y,I,I_{3})$ with $Y$ being hypercharge, $I$ and $I_{3}$
isospin and its third component. Left index $S=(Y_{R}=-N_{c}/3,S,S_{3})$ and
$J=(-Y_{R}=N_{c}/3,S,-S_{3})$ is a state conjugate to $S$. The fact that the
right index runs only over the SU(2) subspace of the SU(3) representation
$\mathcal{R}$ follows from the form of the hedgehog ansatz for the static
soliton field and the Wess-Zumino term \cite{SU3SM,Witten}.
$Q_{S}$ is the charge of state $S$. Note that under the action
of left (flavor) generators $\hat{T}_{\alpha}=-D_{\alpha\beta}^{(8)}%
\hat{J}_{\beta}$ $\psi_{B_{\mathcal{R}}}$ transforms like a tensor in
representation $\mathcal{R}$, while under the right generators
$\hat{J}_{\alpha}$ like a tensor in $\mathcal{R}^{\ast}$ rather than
$\mathcal{R}$.

\begin{figure}[t]
\begin{center}
\includegraphics[height=1.4183in,width=5.3757in]{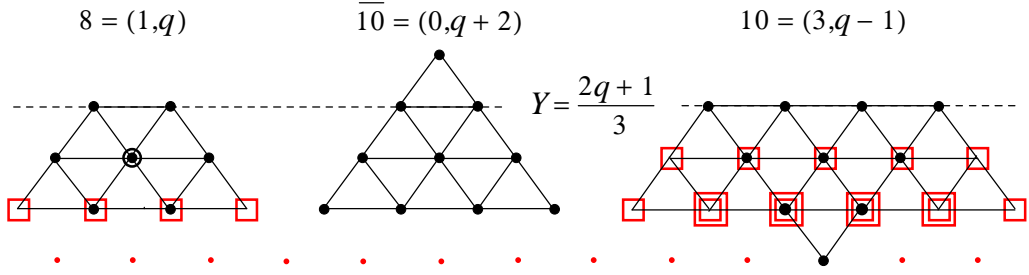}
\end{center}
\caption{Generalizations of octet, antidecuplet and decuplet for large $N_{c}%
$. Bold dots and circles denote physical states, squares and dots at the
bottom denote spurious states which disappear in the limit $q\rightarrow1$.}%
\label{fig:reps}%
\end{figure}

For $N_{c}=3$ $Y_{R}=-1$ and the lowest allowed SU(3) representations are
those of triality zero: $8$, $10$, $\overline{10}$, $27$ etc. In large $N_{c}$
limit, however, these representations are generalized in the following way
\cite{largereps,dulmp} (other possible choices are discussed in Ref.\cite{dul}):
\begin{equation}
"8"=(1,q),\quad"10"=(3,q-1),\quad"\overline{10}"=(0,q+2) \label{reps}%
\end{equation}
where\quad%
\begin{equation}
q=\frac{N_{c}-1}{2}. \label{q}%
\end{equation}

Note that in this notation anti-ten is not a conjugate of
deccuplet, neither is an octet selfadjoint. Therefore we shall
denote a complex conjugate by a
$\ast$:%
\begin{equation}
\mathcal{R}^{\ast}=(p,q)^{\ast}=(q,p). \label{cc}%
\end{equation}
For large representations like (\ref{reps}) physical flavor states have
hypercharge different from the one in the real world, however isospin, charge
and strangeness take the physical values. This is depicted in Fig. 1.

Baryon masses are given in terms of the effective hamiltonian%
\[
M=M_{sol}+\frac{1}{2I_{1}}S(S+1)+\frac{1}{2I_{2}}\left(  C_{2}(\mathcal{R}%
)-S(S+1)-\frac{N_{{c}}^{2}}{12}\right)
\]
where $C_{2}(\mathcal{R})=\left(  p^{2}+q^{2}+pq+3(p+q)\right)  /3$ is the
Casimir operator for representation $\mathcal{R}=(p,q)$ and $M_{sol}%
=\mathcal{O}(N_{c})$ is the classical soliton mass. It is easy to convince
oneself that in the chiral limit nonexotic splittings scale like $\mathcal{O}%
(1 / N_{c})$ while exotic-nonexotic like $\mathcal{O}(1)$, e.g.:%
\begin{align}
M_{\Delta}-M_{N}  &  \sim\mathcal{O}(\frac{1}{N_{c}}),\nonumber\\
M_{\Theta}-M_{N}  &  \sim\mathcal{O}(1). \label{Ncscaling}%
\end{align}
The fact that $\Theta-N$ mass difference is of the order of 1
was used to argue that the rigid-rotor quantization of the chiral
soliton is not valid for exotic states \cite{Cohen}. The discussion
of this point is beyond the scope of the present paper, let us
however note, that arguments have been also given in favor of the
rigid-rotor quantization \cite{DPNc} despite Eq.(\ref{Ncscaling})

\section{Decay width}

The baryon-meson ($\kappa$) coupling operator can be written in terms of the
collective coordinates as \cite{DPP}:%
\begin{equation}
\hat{O}_{\kappa}=-i\frac{3}{2M_{B}}\left[  G_{0}\hat{O}_{\kappa A}^{(0)}%
-G_{1}\hat{O}_{\kappa A}^{(1)}-G_{2}\hat{O}_{\kappa A}^{(2)}\right]  p_{A}.
\label{Op}%
\end{equation}
where%
\begin{equation}
\hat{O}_{\kappa A}^{(0)}=D_{\kappa A}^{(8)},\;\hat{O}_{\kappa A}^{(1)}%
=d_{Abc}\,D_{\kappa b}^{(8)}\hat{S}_{c},\;\hat{O}_{\kappa A}^{(2)}%
=\frac{1}{\sqrt{3}}D_{\kappa8}^{(8)}\hat{S}_{A}. \label{Ops}%
\end{equation}
Here $\hat{S}_{a}$ are generalized spin right  SU(3) generators related
to the known "isospin", $V$-spin and $U$-spin operators in the following way%
\begin{equation}
\hat{I}_{3}=\hat{S}_{3},\;\hat{I}_{\pm}=\hat{S}_{1}\pm i\hat{S}_{2},\;V_{\pm
}=\hat{S}_{4}\pm i\hat{S}_{5},\;U_{\pm}=\hat{S}_{6}\pm i\hat{S}_{7}%
,\;\hat{Y}=\frac{2}{\sqrt{3}}\hat{S}_{8}. \label{gens}%
\end{equation}
Note that these operators act on the right index of the wave
function (\ref{wf}), for which "isospin" is related to the
physical spin. We have adopted here a convention that Greek
indices run over all possible values: $\alpha,\beta,\ldots,$
$\kappa=1\ldots,8$, capital Latin indices over the SU(2) subgroup:
$A,B\ldots=1,2,3$ and small Latin indices $a,b,c,\ldots =4,5,6,7$.
In order to calculate the width for the decay $B\longrightarrow
B^{\prime}+\kappa$ we have to evaluate the matrix element of
$\hat{O}_{\kappa}$ between the baryon wave functions, square it,
average over
initial and sum over final spin and isospin \cite{DPP}:%
\begin{equation}
\overline{\mathcal{M}}_{B}^{2}=\frac{1}{(2I_{B}+1)(2S_{B}+1)}
\sum\limits_{I_{B\,3},S_{B\,3}}\;
\sum\limits_{I_{B^{\prime}\,3},S_{B^{\prime}\,3}}
\left|
\left\langle
B^{\prime},\,S_{B^{\prime}\,3}\left|  \hat{O}_{\kappa}\right|  B,\,S_{B\,3}%
\right\rangle \right|  ^{2} \label{Mav}%
\end{equation}

Coupling constants $G_{i}$ are related to the axial-vector couplings by a
Goldberger-Treiman relation and scale differently with $N_{c}$:%
\begin{equation}
G_{0}\sim N_{c}^{3/2},\quad G_{1},G_{2}\sim N_{c}^{1/2}. \label{G012}%
\end{equation}
Finally, in order to get the width one has to multiply (\ref{Mav}) by the
phase space volume and the final result reads%
\begin{equation}
\Gamma_{B}=\frac{1}{2\pi}\overline{\mathcal{M}}_{B}^{2}\,p \label{width}%
\end{equation}
where%
\begin{equation}
p=\left|  \vec{p}_{\kappa}\right|  =
\frac{\sqrt{(M^{2}-(M^{\prime}+m_{\kappa
})^{2})(M^{2}-(M^{\prime}-m_{\kappa})^{2})}}{2M} \label{mom-p}%
\end{equation}
is the momentum of meson $\kappa$. In Ref.\cite{DPP} Eq.(\ref{width}) was
multiplied
by a ratio of the baryon masses which is important for the numerical results,
which, however, scales as $\mathcal{O}(1)$ with $N_{c}$ and therefore is
irrelevant for further discussion.

\begin{figure}[t]
\begin{center}
\includegraphics[scale=0.7]{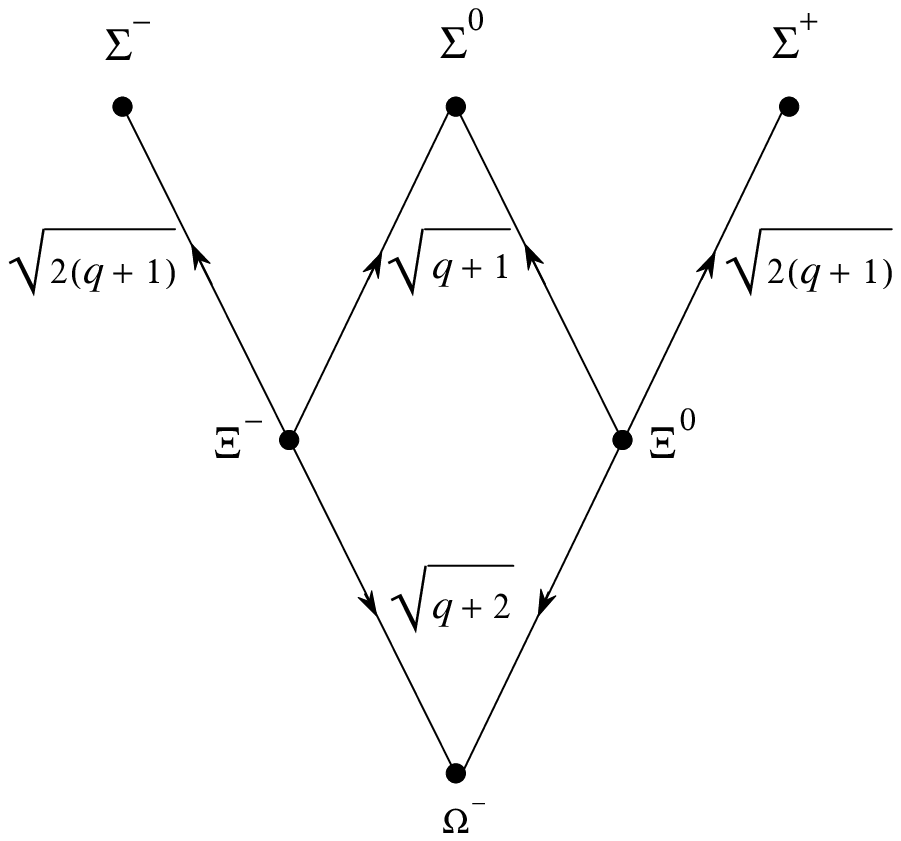}
\includegraphics[scale=0.7,clip=]{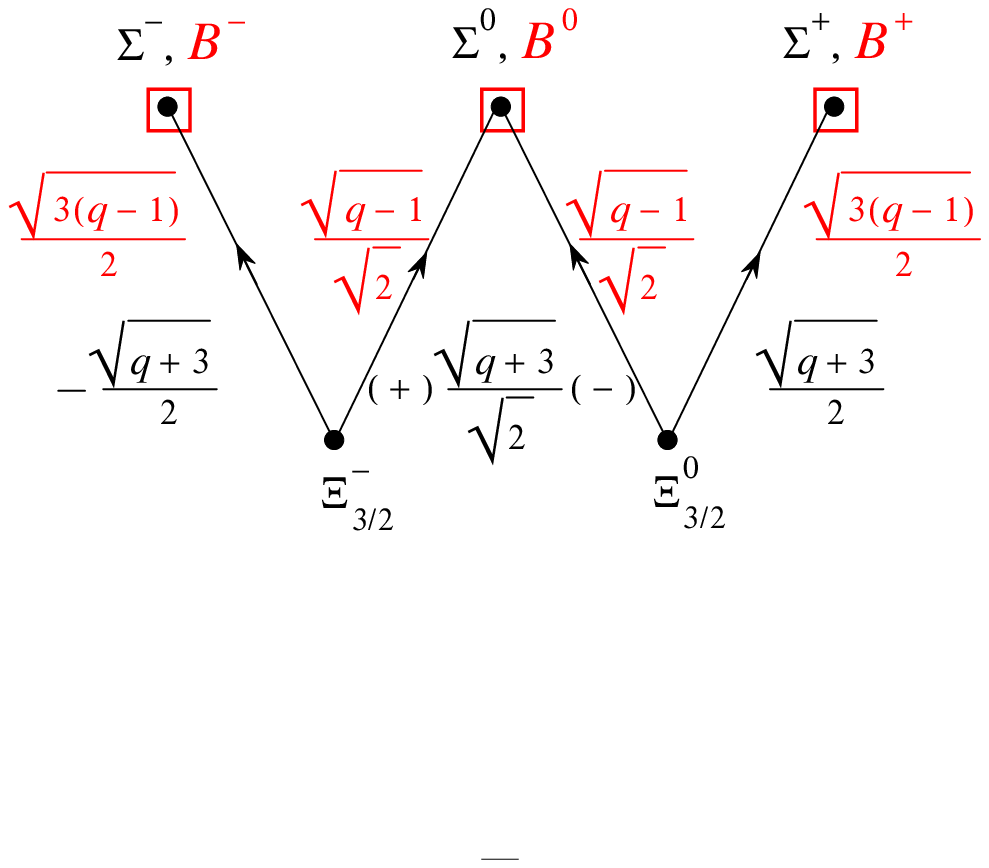}
\end{center}
\caption{Action of $U_{\pm}$ and $V_{\pm}$ operators on spin $S_{3}=\pm1/2$
states belonging to antidecuplet $\mathcal{R}^{\ast}=(q+2,0)$ and decuplet
$\mathcal{R}^{\ast}=(q-1,3)$. In the latter case the upper entries refer to the
transitions to the spurious states $B$. }%
\label{fig2}%
\end{figure}

The action of the $D$ functions entering the collective operators can be
calculated with the help of the SU(3) Clebsch-Gordan coefficients
\cite{DS,WK}:
\begin{align}
&  \mathrm{dim}\;(\mathcal{R}_{3})\,\int dA\,D_{B_{3}J_{3}}^{(\mathcal{R}%
_{3})\ast}{(A)}\,D_{B_{2}J_{2}}^{(\mathcal{R}_{2})}(A)\,D_{B_{1}J_{1}%
}^{(\mathcal{R}_{1})}(A)\nonumber\\
&  =\sum\limits_{\gamma}\left(  \left.
\begin{tabular}
[c]{cc}%
$\mathcal{R}_{1}$ & $\mathcal{R}_{2}$\\
$B_{1}$ & $B_{2}$%
\end{tabular}
\right|
\begin{tabular}
[c]{c}%
$\mathcal{R}_{3}^{\gamma}$\\
$B_{3}$%
\end{tabular}
\right)  \left(  \left.
\begin{tabular}
[c]{cc}%
$\mathcal{R}_{1}$ & $\mathcal{R}_{2}$\\
$J_{1}$ & $J_{2}$%
\end{tabular}
\right|
\begin{tabular}
[c]{c}%
$\mathcal{R}_{3}^{\gamma}$\\
$J_{3}$%
\end{tabular}
 \right)
\end{align}
where $\gamma$ is the degeneracy index. In order to calculate matrix elements
of (\ref{Ops}) between wave functions (\ref{wf}) we shall also use the action
of the operators $V_{\pm}$ and $U_{\pm}$ on the spin states \cite{DS}:%
\begin{equation}%
\begin{array}
[c]{ccccc}%
U_{+} & \nwarrow &  & \nearrow & V_{+}\\
&  &  &  & \\
V_{-} & \swarrow &  & \searrow & U_{-}%
\end{array}
\end{equation}
Note that the spin states belong to $\mathcal{R}^{\ast}$. The relevant
action is depicted in Fig.\ref{fig2}.

%\begin{figure}[t]
%\begin{center}
%\includegraphics[scale=0.7]{UV.eps}
%\end{center}
%\caption{Schematic action of $U_{\pm}$ and $V_{\pm}$ operators on states in a
%given SU(3) representation $\mathcal{R}$. }%
%\label{fig2}%
%\end{figure}

%EndExpansion

Finally we shall need Clebsch-Gordan coefficients for large SU(3)
representations (\ref{reps}). Here we list the Clebsch-Gordan series for the
highest weights in two cases relevant to the present calculation
\cite{largereps,dulmp}:%
\begin{align}
\left|  "10",\Delta^{++}\right\rangle  &  =\sqrt{\frac{q}{q+1}}\left|
8,\pi^{+}\right\rangle \otimes\left|  "8",p\right\rangle -\sqrt{\frac{1}{q+1}%
}\left|  8,K^{+}\right\rangle \otimes\left|  "8",\Sigma^{+}\right\rangle
,\nonumber\\
\left|  "\overline{10}",\Theta^{+}\right\rangle  &  =\sqrt{\frac{1}{2}}\left|
8,K^{0}\right\rangle \otimes\left|  "8",p\right\rangle -\sqrt{\frac{1}{2}%
}\left|  8,K^{+}\right\rangle \otimes\left|  "8",n\right\rangle . \label{HW}%
\end{align}
Remaining Clebsch-Gordan coefficients can be found by applying lowering
operators to (\ref{HW}).

The result of the calculations are as follows (for spin down states and
$\vec{p}=(0,0,p)$):

\begin{itemize}
\item $\Delta\rightarrow\pi N$%
\begin{align}
\left\langle N\left|  \hat{O}_{\pi}\right|  \Delta\right\rangle  &
=-i\frac{3}{M_{N}+M_{\Delta}}\left(  \left.
\begin{tabular}
[c]{cc}%
$8$ & $"8"$\\
$\pi$ & $N$%
\end{tabular}
\ \right|
\begin{tabular}
[c]{c}%
$"10"$\\
$\Delta$%
\end{tabular}
\ \right)  \sqrt{\frac{q+3}{3(q+4)}}\nonumber\\
&  \cdot\left[  G_{0}+\frac{1}{2}G_{1}\right]  \cdot p, \label{Ddcy}%
\end{align}

\item $\Theta^{+}\rightarrow KN$%
\begin{align}
\left\langle N\left|  \hat{O}_{K}\right|  \Theta^{+}\right\rangle  &
=-i\frac{3}{M_{N}+M_{\Theta^{+}}}\left(  \left.
\begin{tabular}
[c]{cc}%
$8$ & $"8"$\\
$K$ & $N$%
\end{tabular}
\ \ \right|
\begin{tabular}
[c]{c}%
$"10"$\\
$\Theta^{+}$%
\end{tabular}
\ \ \right)  \sqrt{\frac{q+1}{2(q+2)(q+4)}}\nonumber\\
&  \cdot\left[  G_{0}-\frac{q+1}{2}G_{1}-\frac{1}{2}G_{2}\right]  \cdot p.
\label{Thdec}%
\end{align}
\end{itemize}

For $\Delta$ decay we get:%
\begin{align}
\overline{\mathcal{M}}_{\Delta}^{2}  &  =\frac{3}{(M_{N}+M_{\Delta})^{2}%
}\frac{q(q+3)}{2(q+1)(q+4)}\left[  G_{0}+\frac{1}{2}G_{1}\right]  ^{2}%
p^{2}\nonumber\\
&  \frac{3}{(M_{N}+M_{\Delta})^{2}}\frac{(N_{c}-1)(N_{c}+5)}{2(N_{c}%
+1)(N_{c}+7)}\left[  G_{0}+\frac{1}{2}G_{1}\right]  ^{2}p^{2}. \label{MDav}%
\end{align}
whereas for $\Theta^{+}$we have:%
\begin{align}
\overline{\mathcal{M}}_{\Theta}^{2}  &  =\frac{3}{(M_{N}+M_{\Theta})^{2}%
}\frac{3(q+1)}{2(q+2)(q+4)}\left[  G_{0}-\frac{q+1}{2}G_{1}-\frac{1}{2}%
G_{2}\right]  ^{2}p^{2}\nonumber\\
&  \frac{3}{(M_{N}+M_{\Theta})^{2}}\frac{3(N_{c}+1)}{(N_{c}+3)(N_{c}%
+7)}\left[  G_{0}-\frac{N_{c}+1}{4}G_{1}-\frac{1}{2}G_{2}\right]  ^{2}p^{2}.
\label{MThav}%
\end{align}

Two important remarks are here in order. First of all, and this is
our main result announced in the Introduction, for $\Theta^+$
decay constant $G_{1}$ is enhanced by a factor of $N_{c}$ and
therefore the second term in Eq.(\ref{MThav}) is of the same order
as $G_{0}$. These two terms cancel against each other yielding
numerically $\Theta^+$ width much smaller than the width of
$\Delta$. This cancellation is therefore consistent with $N_{c}$
counting and justifies the use of nonleading terms in the decay
operator (\ref{Op}).

Secondly, the overall factor in front of $\left[  \ldots\right]
\times p^{2}$ is $\mathcal{O}(1)$ for $\Delta\rightarrow\pi N$ and
$\mathcal{O}(1/N_{c})$ for $\Theta^{+}\rightarrow KN$. This
effect, as can be seen from Es.(\ref{Ddcy},\ref{Thdec}), is
entirely due to the ''spin'' part of the matrix elements $<
B^{\prime} | \hat{O}_{\kappa}|B > $.
Indeed, flavor Clebsch-Gordan coefficients in Es.(\ref{Ddcy}%
,\ref{Thdec}) scale as $\mathcal{O}(1)$ with $N_{c}$, as can be
inferred from Es.(\ref{HW}). This is, however, not a complete
$N_{c}$ dependence, since momentum $p$ also depends on $N_{c}$. We
shall come back to this dependence in a moment.

After multiplying by the phase factors we get:%
\begin{align}
\Gamma_{\Delta}  &  =\frac{3}{2\pi(M_{N}+M_{\Delta})^{2}}\frac{(N_{c}%
-1)(N_{c}+5)}{2(N_{c}+1)(N_{c}+7)}\left[  G_{0}+\frac{1}{2}G_{1}\right]
^{2}p^{3},\nonumber\\
\Gamma_{\Theta}  &  =\frac{3}{2\pi(M_{N}+M_{\Theta})^{2}}\frac{3(N_{c}%
+1)}{(N_{c}+3)(N_{c}+7)}\left[  G_{0}-\frac{N_{c}+1}{4}G_{1}-\frac{1}{2}%
G_{2}\right]  ^{2}p^{3} \label{widths}%
\end{align}
where $p$ is given by Eq.(\ref{mom-p}). In the chiral limit $m_{\kappa}=0$%
\begin{equation}
p=\frac{(M-M^{\prime})(M+M^{\prime})}{2M}.
\end{equation}
Since the difference $M-M^{\prime}$ scales differently with $N_{c}$ for
$\Delta$ and $\Theta^+$ decays (\ref{Ncscaling}):%
\begin{equation}
\Delta\rightarrow\pi N\quad p_{\pi}\sim\mathcal{O}(\frac{1}{N_{c}}),\qquad
\Theta\rightarrow KN\quad p_{K}\sim\mathcal{O}(1) \label{pscale0}%
\end{equation}
and the overall scaling for the widths reads%
\begin{equation}
\Gamma_{\Delta}\sim\mathcal{O}(\frac{1}{N_{c}^{2}}),\qquad\Gamma_{\Theta}%
\sim\mathcal{O}(1). \label{Gscale0}%
\end{equation}

It is interesting to ask at this point how well the $N_c$ scaling
arguments work numerically in Nature. For the mass differences we get
(assuming $M_{\Theta^+}=1540$, $M_{\Xi_{3/2}}=1860$~MeV, which gives
$\overline{M}_{\overline{10}}=1752$~MeV for the average antidecuplet
mass):
\begin{equation}
\overline{M}_{10}-\overline{M}_{8}=234, \qquad
\overline{M}_{\overline{10}}-\overline{M}_{8}=601
\end{equation}
(in MeV) which is in a reasonable agreement with an expected factor of
$N_c=3$, see Eq.(\ref{Ncscaling}). As far as the phase-space factor is
concerned, the physical value of the meson momentum reads (in MeV)
\begin{equation}
p_{\pi}=225, \qquad p_{K}=268
\end{equation}
and it is hard to argue that the scaling of Eq.(\ref{pscale0}) really holds.
Formally for $m_{\kappa}\neq0$ meson momentum $p$ scales in both cases
\cite{foot3} as $\mathcal{O}(1)$.
In that case:%
\begin{equation}
\Gamma_{\Delta}\sim\mathcal{O}(N_{c}),\qquad\Gamma_{\Theta}\sim\mathcal{O}(1)
\label{Gscalem}%
\end{equation}
which would explain parametrically the narrowness of $\Theta^{+}$ with respect
to $\Delta$.

\section{Nonrelativistic limit}

Coupling constants $G_{0,1,2}$ are related to the axial couplings through
Goldberger-Treiman relation. On the other hand we know explicit model formulae
for these couplings \cite{Blotz,paradox}:%
\begin{equation}
G_{0}\sim A_{0}-\frac{A_{1}^{(-)}}{I_{1}^{(+)}},\;G_{1}\sim2\frac{A_{2}^{(+)}%
}{I_{2}^{(+)}},\;G_{2}\sim2\frac{A_{1}^{(+)}}{I_{1}^{(+)}}%
\end{equation}
up to the same proportionality factor of the order of $M_{B}/F_{\pi}%
\sim\mathcal{O}(\sqrt{N_{c}})$. Explicit formulae and numerical values of
the inertia parameters $A$ and $I$ can be found in Ref.\cite{Blotz}.
If in the $\chi$QSM one artificially sets
the soliton size $r_{0}\rightarrow0$, then the model reduces to the free
valence quarks which, however, ''remember'' the soliton structure \cite{limit}.
In this limit, many quantities, like the axial-vector couplings, are given as ratios
of the group-theoretical factors \cite{paradox}:%
\begin{equation}
A_{0}\rightarrow-N_{c}\ ,~~~~\frac{A_{1}^{(+)}}{I_{1}^{(+)}}\rightarrow
-1\ ,~~~~\frac{A_{1}^{(-)}}{I_{1}^{(+)}}\rightarrow2\ ,~~~~\frac{A_{2}^{(+)}%
}{I_{2}^{(+)}}\rightarrow-2\;.
\end{equation}
With these values we get that the nucleon axial coupling \cite{paradox,limit}%
\begin{equation}
g_{A}\longrightarrow\frac{N_{c}+2}{3}=\frac{5}{3}%
\end{equation}
which is the well known naive quark model result \cite{Karl}.

For the antidecuplet decay strength we get:%
\begin{align}
G_{\overline{10}} &=  G_{0}-\frac{N_{c}+1}{4}G_{1}-\frac{1}{2}%
G_{2} \nonumber \\
 &  \sim\left(  A_{0}-\frac{A_{1}^{(-)}}{I_{1}^{(+)}%
}\right)  -\frac{N_{c}+1}{2}\frac{A_{2}^{(+)}}{I_{2}^{(+)}}-\frac{A_{1}^{(+)}%
}{I_{1}^{(+)}}\nonumber\\
&  =\left(  A_{0}-\frac{N_{c}}{2}\frac{A_{2}^{(+)}}{I_{2}^{(+)}}\right)
+\left(  -\frac{A_{1}^{(-)}}{I_{1}^{(+)}}-\frac{1}{2}\frac{A_{2}^{(+)}}%
{I_{2}^{(+)}}-\frac{A_{1}^{(+)}}{I_{1}^{(+)}}\right) \nonumber\\
&  \longrightarrow\left(  -N_{c}+\frac{N_{c}}{2}2\right)  +(-2+\frac{1}%
{2}2+1)=0. \label{G10bar}%
\end{align}
We see that the cancellation is exact in each order in $N_{c}$~\cite{foot2}.

Let us speculate that the cancellation of the leading order terms in
(\ref{G10bar}) is exact. Then $G_{\overline{10}}\sim\mathcal{O}(\sqrt{N_{c}})$
while $G_{10}\sim\mathcal{O}(N_{c}^{3/2})$. In this case we would get%
\begin{align}
\Gamma_{\Delta}  &  \sim\,\mathcal{O}(N_{c})\,\times\,p^{3}\rightarrow
\left\{
\begin{array}
[c]{ccc}%
\mathcal{O}(\frac{1}{N_{c}^{2}}) &  & m_{\kappa}=0\\
\mathcal{O}(N_{c}) &  & m_{\kappa}\neq0
\end{array}
\right. \nonumber\\
\Gamma_{\Theta}  &  \sim\mathcal{O}(\frac{1}{N_{c}^{2}})\times p^{3}%
\rightarrow\left\{
\begin{array}
[c]{ccc}%
\mathcal{O}(\frac{1}{N_{c}^{2}}) &  & m_{\kappa}=0\\
\mathcal{O}(\frac{1}{N_{c}^{2}}) &  & m_{\kappa}\neq0
\end{array}
\right.
\end{align}
which would mean that in the chiral limit both decay widths scale as
$\mathcal{O}(\frac{1}{N_{c}^{2}})$ while for $m_{\kappa}\neq0$ $\Theta^{+}$
decay is damped by a factor $\mathcal{O}(N_{c}^{3})\,$with respect to $\Delta$ .

\section{Summary}

Our primary goal was to show that the cancellation which takes
place in the case of the $\Theta^{+}$ width is consistent with the
$N_{c}$ counting. Indeed, by employing correct generalizations of
standard SU(3) representations for arbitrary number of colors, we
have shown that there is additional $N_{c}$ enhancement of the
constant $G_{1}$ which is formally one power of $N_{c}$ less than
$G_{0}$. This enhancement comes entirely from the spin part of the
matrix elements $< B^{\prime} | \hat{O}_{\kappa}|B > $ and carries
over to the decays of all particles in antidecuplet.

We have also found that there is $\mathcal{O}(1/N_{c})$
suppression factor in the $\Theta^{+}$ width with respect to
$\Delta$, coming from the same source, namely from the $N_{c}$
dependence of the SU(3)$_{\text{flavor}}$ Glebsch-Gordan
coefficients. Unfortunately, this suppression is ''undone'' by the
phase space factor $p^{3}$, which in the chiral limit scales
differently for $\Theta^{+}$ and $\Delta$ decays (\ref{pscale0}).
If we assume that meson masses are nonzero, then the suppression
survives (\ref{Gscalem}). This kind of ''noncommutativity'' of the
chiral limit and large $N_{c}$ expansion is well known and there
are many examples where it creates problems.

Finally, we have shown that in the nonrelativistic quark model limit,
i.e. in the limit
where we artificially squeeze the soliton, the cancellation in the decay
strength for $\Theta^{+}$ is exact and occurs independently at each order of
$N_{c}$. In this limit $\Theta^+$ width vanishes identically.

\section*{Acknowledgments}

The author would like to thank Dmitri Diakonov for stimulating
discussion and Thomas Cohen for useful remarks. This work has been
partially supported by the Polish State Committee for Scientific
Research under grant 2 P03B 043 24 and it has been authored under
Contract No. DE-AC02-98CH10886 with the U. S. Department of
Energy.

%%%%%%%%%%%%%%%%%%%%%%%%%%%%%%%%%%%%%%%%%%%%%%%%%%%%%%%%%%%%%%%%%%%%%%%%%%%%%


\begin{thebibliography}{99}
\bibitem{foot3}
In the case of $\Delta $ decay $p_{\pi}$ is imaginary
since, strictly speaking, in the large $N_c$ limit $M_{\Delta }=M_N$
and the decay does not occur.
\end{thebibliography}
\end{document}